# Soil Salinity Frequency-Dependent Prediction Model Using Electrical Conductivity Spectroscopy Measurement

[1] Javad Jafaryahya, [1]Rasool Keshavarz, [2]Tarou Kikuchi, [1]Negin Shariati

*Abstract*— Soil salinity is a critical factor influencing agricultural productivity and environmental sustainability, requiring precise monitoring tools. This paper focuses on developing a frequency-dependent model to predict soil salinity based on electrical conductivity (EC) and volumetric water content (VWC). A dataset of 40 soil samples with varying levels of salinity and moisture, consisting of two soil types (sandy and clayey), was experimentally measured for EC in the frequency range of 10 to 295 MHz using electrical conductivity spectroscopy (ECS) measurement with the DAK-VNA (Dielectric Assessment Kit - Vector Network Analyzer) system. A new, more comprehensive frequency-dependent model is proposed, surpassing previous models that lacked frequency considerations. This modelling approach was conducted in stages: initially, a frequency-independent model for electrical conductivity as a function of salinity and moisture was developed. Next, a frequency-dependent model was introduced. Finally, a comparison between pure sandy soil and a sandy-clay mixture led to the final model, which also incorporates effective porosity. The results of the proposed model, comparing measured and predicted values, provide a robust approach to accurately predict soil salinity. Findings demonstrate that the model can enhance salinity prediction accuracy, extending its applicability beyond agriculture to geological and hydrological applications in real-world scenarios.

## I. Introduction

Soil salinity plays a critical role in agriculture, environmental sustainability, and soil resource management. While certain salts, such as potassium chloride, can enhance soil fertility and promote plant growth by providing essential nutrients, excessive salt levels, particularly from sodium chloride, can have detrimental effects. High soil salinity disrupts plant water uptake, causes ion toxicity, and leads to nutrient imbalance, thereby deteriorating soil structure and reducing agricultural productivity [1]. The accumulation of soluble salts like potassium, sodium, magnesium, and calcium in the soil water further compounds these effects, influencing soil properties and posing a risk of reduced crop yields or even crop failure. Maintaining optimal salinity levels is thus essential for preserving soil health and supporting sustainable agricultural practices. [2, 3].

Several methods are available for detecting and measuring soil salinity, ranging from laboratory-based techniques to field-based and remote sensing approaches. Laboratory methods often involve measuring the electrical conductivity (EC) of soil extracts, a widely used metric where higher EC values indicate increased salinity levels[4]. Other lab-based techniques include chemical analysis of soil extracts to identify specific ions, such as sodium, potassium, calcium, and magnesium, which contribute to salinity, as well as ion-specific spectroscopy methods like Atomic Absorption Spectroscopy (AAS) and Inductively Coupled Plasma (ICP) spectroscopy [5]. Field-based approaches utilize portable EC sensors for rapid, in-situ salinity measurements, providing instant feedback on soil conditions. Capacitive sensing methods, Time Domain Reflectometry (TDR), and Frequency Domain Reflectometry (FDR) are also commonly used to determine soil conductivity and analyze soil moisture and salinity based on the electrical properties of the soil [6-12]. Additionally, Near-Infrared (NIR) spectroscopy has gained attention as a non-invasive technique to estimate soil salinity by analysing light reflectance off the soil surface. In large-scale agricultural and environmental monitoring, remote sensing and satellite imagery offer valuable insights into soil salinity distribution by analysing spectral data over vast areas, particularly for salinity mapping in arid and semi-arid regions [13-15]. Each of these methods offers unique advantages and limitations based on cost, accuracy, and scale of application.

Soil electrical properties play a crucial role in assessing salinity and water content, both essential factors for effective agricultural management and productivity [16]. Parameters such as EC and permittivity offer valuable insights into soil volumetric water content (VWC) and salinity levels [7, 9, 17-21]. In particular, apparent soil electrical conductivity ($EC_a$) is widely applied to track long-term changes in soil, including salinization, which directly influences crop yields [22-24]. The agricultural use of $EC_a$ was pioneered by Rhoades et al. [25], who demonstrated its effectiveness for estimating soil salinity in various crop settings. The relationship between $EC_a$ and soil characteristics—such as texture, VWC, and salinity—is, however, complex and context-specific, varying considerably with field conditions. Building on this, Gupta and Hanks [26] explored how soil water and salt levels influence $EC_a$, showing that $EC_a$ rises with increases in both parameters. Numerous studies have since affirmed that $EC_a$ readings are affected by multiple factors, including soil texture, water content, and salt concentration [27-29].

[1]The authors are with the Radio Frequency and Communication Technologies (RFCT) Research Laboratory, University of Technology Sydney, Ultimo, NSW 2007, Australia (e-mail: javad.jafaryahya@student.uts.edu.au; negin.shariati@uts.edu.au)

[2] NTT Communications Corporation.

Developing models to describe soil properties is of special importance for advancing digital agriculture [30]. This research focuses on developing a frequency-dependent model to predict soil salinity using soil electrical properties, such as EC and VWC. Previous models in this domain have typically been frequency-independent, with complex requirements for estimating salinity parameters. In contrast, the proposed model leverages characteristics that can be measured with standard commercial sensors. Given that most sensors operate at specific frequencies, a frequency-dependent model has been designed to improve the accuracy of soil salinity predictions. To construct this model, a series of soil samples with varying salinity levels (0-2 g/kg) and VWC percentages (1-15%) was created, and their EC was measured across a frequency range of 10 to 295 MHz using the DAK and VNA (Dielectric Assessment Kit - Vector Network Analyzer) system. Initially, the model was developed using sandy soil samples alone, but in the next stage, a mix of sandy and clay soil samples was introduced to assess the influence of effective porosity. The modelling process was conducted in phases, beginning with a frequency-independent model that related EC to salinity and VWC. Subsequently, a frequency factor covering the 10-295 MHz range was incorporated, and finally, the influence of effective porosity was integrated into the final model. This approach resulted in a model that is more versatile and adaptable than previous models. Optimization techniques were employed to identify the best-fit functions, making the model applicable not only to agriculture but also to fields such as geology and hydrology.

The steps for designing a soil salinity prediction model based on EC are shown in Fig. 1. The first step is data collection, which involves creating a dataset with soil samples at varying salinity and moisture levels and measuring EC over a defined frequency range (e.g., 10–295 MHz). After pre-processing to ensure measurement accuracy, effective features impacting salinity, such as moisture content, soil type, and frequency response, are identified and analysed. These insights guide the design of a parametric model, where the behaviour of each parameter and variable helps inform the model structure, and the best-fit model is selected through error optimization. Initially, a frequency-independent model is proposed, followed by an addition of frequency dependency in the next phase. Finally, the model undergoes evaluation and validation to ensure reliability.

The paper is organized as follows: Section II reviews well-known models for predicting soil salinity. Section III introduces the dataset and presents the measured data. Section IV introduces the proposed model, while Section V discusses the conclusions and findings.

## II. REVIEW OF KEY SOIL SALINITY MODELS

In soil science, different forms of EC are used to understand how electrical currents move through soils, each offering unique insights into various soil properties. EC in soils provides crucial information on moisture levels, salinity, and ion exchange, which are fundamental for agricultural productivity and environmental health [2, 3].

Apparent electrical conductivity ($EC_a$) is the most measured form, reflecting the overall conductivity of the soil. It is influenced by multiple factors such as salinity, moisture, soil texture, and temperature. $EC_a$ represents a combination of bulk liquid-phase conductivity (associated with dissolved salts) and surface conductivity (due to adsorbed ions on soil particle surfaces), making it particularly useful for understanding spatial variability in soil properties, especially in precision agriculture [22-24].

Bulk electrical conductivity ($EC_b$) and surface conductivity ($EC_s$) are fundamental components of $EC_a$ that illustrate how electrical currents pass through different parts of the soil. $EC_b$ represents the conductivity arising from dissolved salts like sodium, potassium, and other ions freely present in the liquid phase of the soil, filling the spaces between soil particles. $EC_b$ is significantly influenced by the VWC and essentially measures the soil's capacity to conduct electricity through the unbound liquid phase of the soil matrix[4, 31]. In contrast, $EC_s$ measures conductivity that occurs at the solid-liquid interface. This type of conductivity is particularly significant in soils with high levels of clay or organic matter, where exchangeable ions are held on the surfaces of soil particles rather than being freely dissolved. These ions are adsorbed onto the surfaces of clay minerals and organic materials, and $EC_s$ is therefore highly influenced by the soil's texture and mineral composition. This makes $EC_s$ particularly relevant in soils with a high proportion of fine particles, such as clay [32].

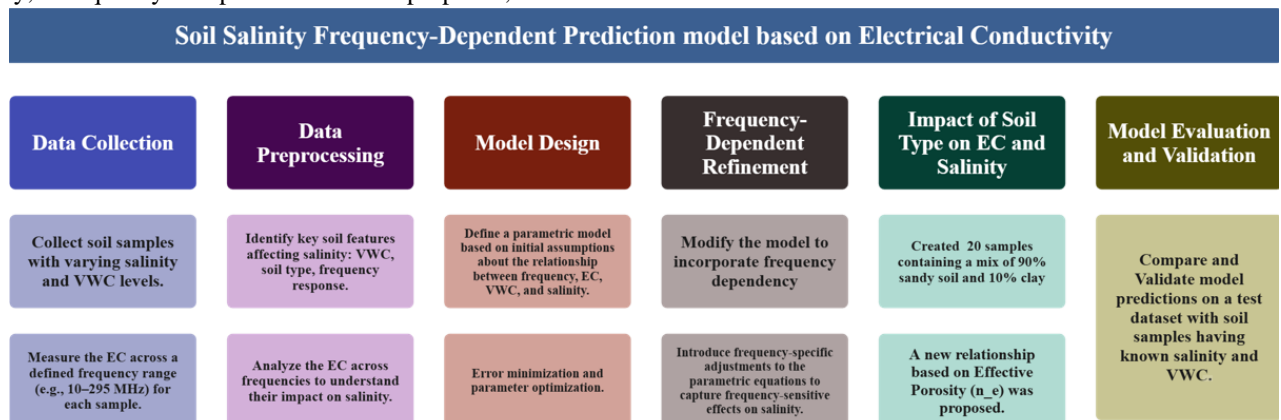

Fig. 1. The steps for designing a soil salinity prediction model based on EC

The bulk soil electrical conductivity ($EC_a$) can be conceptualized as resulting from two parallel conducting pathways. The first pathway is the bulk liquid-phase conductivity ($EC_b$), which is associated with free salts dissolved in the water occupying the soil pores. The second pathway is the bulk surface conductivity ($EC_s$), which is linked to the exchangeable ions at the solid-liquid interface, primarily on the surfaces of clay particles and organic matter. This relationship can be represented by the Equation (1) [31]:

$$EC_a = EC_b + EC_s \tag{1}$$

Water Conductivity ($EC_w$) is the electrical conductivity specifically of the pore water in soil, containing dissolved salts such as sodium, potassium, calcium, and magnesium. This parameter is often what is measured by commercial soil sensors as a standard indicator of soil salinity. $EC_w$ is important because it directly represents the salinity of the soil water, which plays a major role in determining the ability of the soil to conduct electricity through its liquid phase. It is a straightforward and commonly used measure that reflects how well ions are moving through the water in the soil. Since $EC_w$ directly correlates with the concentration of dissolved salts, it can be used to understand the effect of salinity on plant growth and nutrient uptake, making it a critical value for agricultural management and environmental assessments. The bulk liquid-phase conductivity ($EC_b$) is assumed to depend linearly on the electrical conductivity of the soil water ($EC_w$). Moreover, the ability of the soil to conduct electrical current depends on the fraction of the total cross-sectional area occupied by the liquid phase. This means that the apparent electrical conductivity ($EC_a$) in Equation (1) can be rewritten as Equation (2), considering the influence of the VWC ($\theta$). This reformulation provides a clearer representation of how soil moisture and salinity affect conductivity, allowing for a more accurate model of soil electrical behaviour in variable environmental conditions [31].

$$EC_a = EC_w \theta T + EC_s \tag{2}$$

Developing a universal parametric model that can accurately predict soil salinity, considering the complex interplay of factors such as soil type, texture, density, and mineral content, is practically impossible. However, several models have been developed over the years to address specific conditions and soil types [22, 28, 33-35]. One of the most well-known models, Archie's Law, was initially developed in 1942 to estimate water saturation in hydrocarbon reservoirs and has since become a fundamental tool in soil science for estimating soil salinity [16]. This model relates the electrical resistivity or conductivity of a porous medium to its VWC and salinity, providing a simple method for estimating soil salinity levels, as shown in Equation (3).

$$EC_a = a \times \sigma_w \times \varphi^m \tag{3}$$

Where $EC_a$ is the apparent electrical conductivity, a is an empirical constant, $\sigma_w$ represents the conductivity of the pore water, $\varphi$ is the porosity, and m is the cementation factor that varies with soil type and structure. The parameter m is the cementation exponent, which varies depending on the texture and structure of the material. For well-sorted sands, m usually ranges between 1.8 and 2.0, but it can vary for different types of soil or rock. Archie's Law assumes of uniformity in soil structure, making it highly effective for soils with homogeneous texture and composition; However, this limits its accuracy in heterogeneous soils where complex interactions occur between soil particles, salinity, and moisture. In subsequent research, multiple versions of Archie's Law were proposed, attempting to account for these complexities to a greater extent [36] [37, 38].

Saturated Soil Extract Conductivity ($EC_e$) is another key concept in understanding soil salinity. $EC_e$ represents the electrical conductivity of the saturated soil extract, which is a measure of the concentration of soluble salts in a soil solution that has reached its saturation point. $EC_e$ is a standardized way to assess soil salinity, as it measures the conductivity of the soil water under controlled conditions, providing a consistent reference point across different types of soils [28]. Unlike $EC_a$, which measures the total electrical conductivity of soil in its natural, undisturbed state, $EC_e$ focuses specifically on the dissolved salts in a completely saturated solution. This makes $EC_e$ a direct and precise indicator of the salinity stress that plants might experience, providing valuable information for agriculture, especially in saline-prone environments. Based on this approach, Rhoades et al [35] developed a method for estimating soil salinity that dynamically adjusts based on an assumed soil moisture function. The method uses an adaptive VWC function that allows the model to adjust to the existing soil moisture, providing a more reliable prediction of salinity over different environments. This dynamic adjustment makes it effective for real-world agricultural applications where maintaining consistent water levels is challenging. The primary equation used in their approach is given in Equation (4):

$$EC_w = \frac{EC_e \times \rho_b \times S_p}{100 \theta_w} \tag{4}$$

Where $EC_w$ represents the average electrical conductivity of the soil water, $EC_e$ is the electrical conductivity of the saturated soil extract, $\rho_b$ is the bulk density of the soil, $S_p$ stands for the soil saturation percentage, and $\theta_w$ is the VWC. Specifically, by examining the variables in Equation (4), several conclusions can be drawn as follows:

- Linear relationship between $EC_e$ and $EC_w$: Soil salinity ($EC_e$) and the electrical conductivity of soil water ($EC_w$) have a direct linear relationship. As salinity increases, the presence of more dissolved salts (ions) leads to higher conductivity in the soil. This relationship is linear, meaning that if the salinity doubles, the electrical conductivity will also double, assuming other variables remain constant.
- Effect of $\rho_b$ and $S_p$: Bulk density ($\rho_b$) and the saturation percentage ($S_p$) appear in the numerator, suggesting that denser soils with more connected pore spaces conduct electricity more efficiently. Therefore, both have a positive impact on soil conductivity.

It is important to note that in soil, an increase in VWC generally results in a higher degree of saturation, allowing more ions to move freely, thereby increasing conductivity. Hence, while the equation suggests an inverse relationship, it oversimplifies the complex interaction between VWC, salinity, and soil conductivity. More water typically increases the number of conductive pathways, making the system more conductive until saturation is reached [39]. This discrepancy highlights the need to consider ion mobility, soil texture, and water-filled pore connectivity, all of which play a crucial role in determining how VWC affects conductivity under real-world conditions. In general, changes in one factor can affect others, and the combination of factors ultimately determines salinity levels [40]. For example, when discussing the degree of saturation, $S_p$ can be interpreted as the ratio of the current VWC (θ) to the effective soil porosity ($n_e$)[41].

$$S_p = \frac{\theta}{n_e} \quad (5)$$

Now, if Equation (5) is substituted into Equation (4), Equation (6) is obtained for $EC_w$ in terms of θ and $n_e$:

$$EC_w = \frac{EC_e \times \rho_b}{100 \times n_e} \quad (6)$$

This equation shows that the soil's electrical conductivity ($EC_w$) is directly proportional to the electrical conductivity of the pore water ($EC_e$) and the soil's bulk density ($\rho_b$), and inversely proportional to the soil's effective porosity ($n_e$). However, this doesn't mean VWC has no effect in reality; rather, it suggests that the dependency on VWC is already embedded in the relationship between $EC_e$, $\rho_b$, and $n_e$.

Mu et al introduced [42] the ECWS model, a novel approach to measuring soil salinity content (SSC) using soil EC and moisture content (WS). The ECWS model is grounded in the relationship between soil bulk conductivity ($EC_a$), soil solution conductivity $EC_w$, VWC($\theta_c$), and soil salinity content (SSC). The model integrates two key coefficients, $\rho_a$ is the conversion coefficient for soil salinity and conductivity, and $\rho_w$ is for soil extract conductivity. Through the ECWS model, soil salt content is expressed as Equation (7):

$$SSC = \rho_a EC_a + \rho_w \theta_c EC_w + b \quad (7)$$

where b represents initial soil salt content. Experimental validation showed strong correlation between the model's predictions and actual SSC for soils treated with different salts (e.g., NaCl, K2SO4)[42].

### III. Procedure for Measuring EC in the 10-295 MHz

In this research, a soil dataset with varying salinity and moisture levels was purposefully created to investigate the impact of VWC and salinity on soil samples. In this dataset, salinity and VWC were linearly added to pure sandy soil, resulting in 20 different soil types. As shown in TABLE I and Fig. 2, the potassium content in the soil samples varied from 0 to 2 grams per kilogram, while the VWC ranged from 0% to 15%. The soil samples were prepared by mixing pure sandy soil with a potassium chloride solution, which acted as the salinity agent. Potassium chloride, composed of potassium and chlorine, is commonly used as a salt substitute, making it suitable for this study. It appears as a transparent powder with a grain size of 0.02 mm. To add this powder to the soil, a salt solution was first prepared with the amount of water corresponding to the desired VWC percentage and then mixed thoroughly with the soil using a mixer. Soil salinity measurements were carried out with precision, following established procedures and protocols for measuring VWC and potassium levels in different soil types. Before each experiment, all measuring equipment was carefully calibrated to ensure accuracy and reliability. To ensure a fair comparison of data, consistent test conditions were maintained throughout all experiments. All experiments were conducted in a controlled environment with a constant temperature of 23°C, maintained by a temperature control system. This approach ensured that temperature did not affect the experimental results. To prevent water evaporation from the soil samples during the experiments, they were stored in sealed containers, as shown in Fig. 2. Given the potential influence of time on the chemical interactions between potassium and soil, all measurements were conducted promptly to maintain consistent environmental conditions.

For EC measurements over a range of frequencies, the DAK and VNA system was used, as shown in Fig. 2. The DAK and VNA system provide a versatile solution that is used in a wide range of industries, such as electronics, food, chemistry, and medicine[1]. The measurement process involves connecting the probe to a VNA to determine the complex reflection coefficient (S11) at the probe tip. The DAK software then processes the S11 data to calculate the material's complex permittivity (both real and imaginary components) as well as its conductivity. This system includes three different probes: DAK-12 (4 MHz - 3 GHz), DAK-3.5 (200 MHz - 20 GHz), and DAK-1.2E (5 - 67 GHz). For measurements in the range of 10-295 MHz, the model used in this research was the DAKS-12 (4 MHz – 3 GHz), which integrates Copper Mountain Technologies' R60 vector reflectometer[2].

TABLE I. Specifications of 20 Different Soil Samples with Varying Salinity Levels and VWC%

| Num | 1 | 2 | 3 | 4 | 5 | 6 | 7 | 8 | 9 | 10 |
|---|---|---|---|---|---|---|---|---|---|---|
| VWC% | 1 | 1 | 1 | 1 | 1 | 5 | 5 | 5 | 5 | 5 |
| K(g/kg) | 0 | 0.5 | 1 | 1.5 | 2 | 0 | 0.5 | 1 | 1.5 | 2 |
| Num | 11 | 12 | 13 | 14 | 15 | 16 | 17 | 18 | 19 | 20 |
| VWC% | 10 | 10 | 10 | 10 | 10 | 15 | 15 | 15 | 15 | 15 |
| K(g/kg) | 0 | 0.5 | 1 | 1.5 | 2 | 0 | 0.5 | 1 | 1.5 | 2 |

---

[1] https://speag.swiss/zh_cmn/products-zh-cmn/dak-zh-cmn/dielectric-measurements-2-zh-cmn/

[2] https://coppermountaintech.com/vna/r60-1-port/

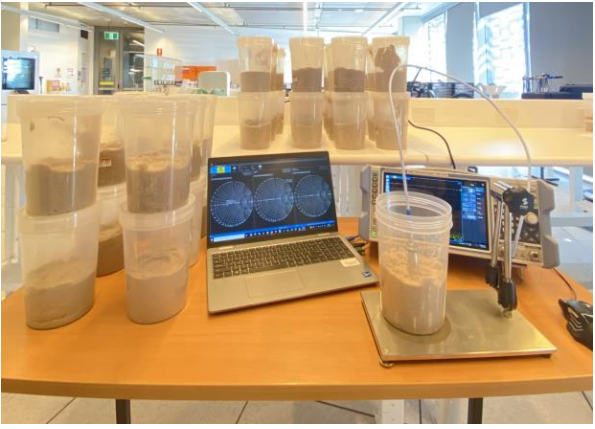

Fig. 2. Soil samples for EC measurement, and the DAK and VNA system

The measurements conducted using the DAK and VNA system as depicted in Fig. 3. These measurements were taken at 58 different frequencies, ranging from 10 MHz to 295 MHz, with a resolution of 5 MHz ([10, 15, 20, ..., 290, 295]). Fig. 3(a) illustrates the electrical conductivity (EC) as a function of frequency (10-295 MHz) for two VWC levels (5% and 1%) across five different potassium levels (0, 0.5, 1, 1.5, 2 g/kg). Additionally, Fig. 3(b) presents data for VWC levels of 10% and 15% with the same potassium levels. As observed, the EC increases with rising salinity and VWC. Furthermore, variations in EC are evident across different frequencies.

The EC measured by the DAK system primarily represents the bulk electrical conductivity ($EC_b$) of the soil, which encompasses both the liquid phase, comprising dissolved salts, and interactions occurring at the solid-liquid interface. This measurement is closely tied to $EC_w$ (water conductivity) since the DAK system evaluates the real part of permittivity and conductivity, both of which are influenced by the ions dissolved in soil water ($EC_w$) as well as other factors such as soil texture and structure. While DAK provides a broader $EC_b$ measurement, it inherently reflects $EC_w$, given that free ions in the water phase significantly impact the overall conductivity. In this study, the electrical conductivity measured by the DAK system will simply be denoted as EC.

By carefully examining Fig. 3 regarding the influence of varying salt level and VWC, it can be concluded that the changes in salt content and VWC are not completely independent in their effects on EC. When one parameter changes (while keeping the other constant), different behaviors in EC are observed at varying levels, suggesting interdependence between these two factors.

These measurements were conducted across a wide frequency range, while the existing models discussed in Section II are frequency independent. As observed, although there are variations in EC values at different frequencies, the average EC over the entire frequency spectrum can be used as an indicator of EC for each soil sample. Fig. 4 shows the averaged EC across the full frequency range for 16 soil samples in relation to VWC. Since the four soil samples with varying VWC but zero potassium level do not influence the modeling, they will be excluded from further analysis in this study.

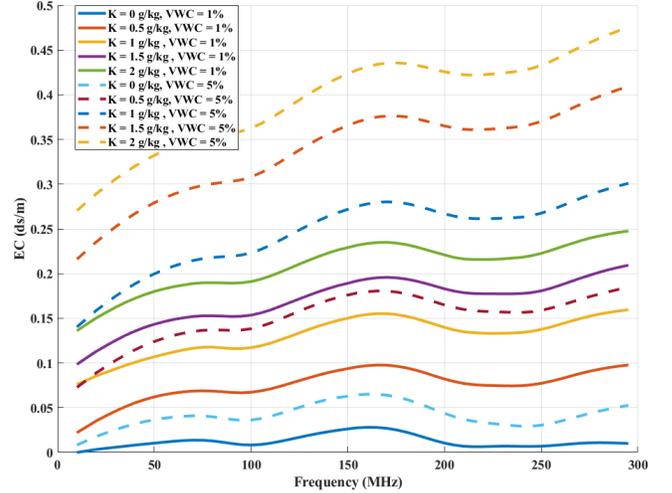

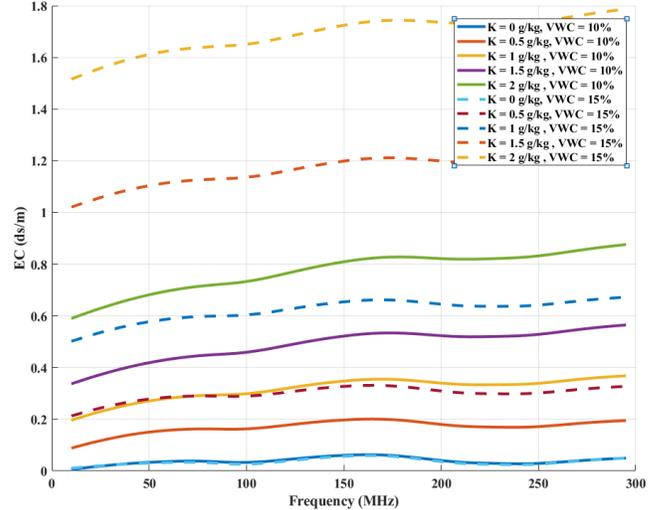

Fig. 3. Measurement of DAK and VNA system, (a) EC as a function of frequency (10-295 MHz) (a) for VWC levels of 1% and 5%, (b) for VWC levels of 10% and 15% across 5 potassium levels (0, 0.5, 1, 1.5, 2 g/kg).

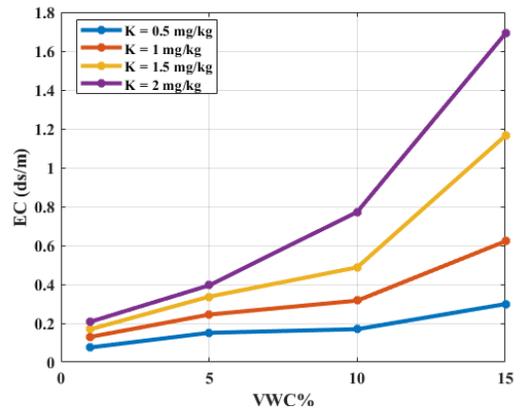

Fig. 4. Averaged EC across the full frequency range of 16 soil samples with VWC greater than 1%.

## IV. Proposed Model for Salinity

In this section, the proposed model is introduced and completed in three stages. First, a frequency-independent model for EC is developed based on salinity and VWC. Since the EC measurements were taken over a wide frequency range, the average conductivity across all frequencies is used as the representative measurement at this stage. Next, to create a frequency-dependent model, a frequency impact factor within the 10-295 MHz range is incorporated as a multiplicative function in the model. Finally, with the addition of 20 new soil samples made from a mix of sand and clay, the influence of effective porosity is integrated as a coefficient, resulting in the final model.

*A. Frequency-independent model for salinity*

For the proposed model, if the effects are considered independently, as in Archie's Law several independent functions can be defined, each accounting for a different factor. These functions will multiply together to give the overall result. The first function will depend on salinity, the second on VWC, and the third on soil texture, as shown in Equation (8).

$$EC_w(S, \theta_w, n_e) = f_1(S) f_2(\theta_w) f_3(n_e) \quad (8)$$

The variable S for salinity is defined within the range [0, 2] g/kg. Additionally, based on the measurements in Section 3, the VWC% ($\theta_w$) varies from 1% to 15%, as shown in Equation (9).

$$S \sim [0,2]\frac{g}{kg}, \quad \theta_w \sim [1,15]\% \quad (9)$$

By expressing each variable as an independent function, their contributions can be separated, allowing their impact on EC to be analyzed. However, upon analyzing Fig. 3, it becomes evident that the assumption of complete independence between VWC and salinity on EC is not entirely accurate given the observed variations. These observations indicate that EC does not change in a perfectly linear manner with salinity changes (assuming constant VWC). The assumption of linearity is merely an approximation. Therefore, Equation (8) needs to be rewritten to account for these interdependencies, resulting in the revised relationship shown in Equation (10).

$$EC_w(S, \theta_w, n_e) = f_1(S, \theta_w) f_2(n_e) \quad (10)$$

After conducting numerous experiments and analyzing the type of changes and dependencies between these two parameters, two types of functions have been proposed for $f_1$. The first one is based on a power function, which is presented in Equation (11) and TABLE II, and the corresponding results are displayed in Fig. 5. Since $EC_e$ represents soil salinity and is directly proportional to the amount of potassium in the soil, in this study, the potassium level is assumed to be equivalent to the salinity or $EC_e$, with a proportionality constant.

$$EC_w(S, \theta_w, n_e) = a_1 S^{(a_2 \theta_w + a_3 S + a_4)} \theta_w^{(a_5 \theta_w + a_6 S + a_7)} f_2(n_e) \quad (11)$$

TABLE II. Coefficients of Equation (11).

| C | $a_1$ | $a_2$ | $a_3$ | $a_4$ | $a_5$ | $a_6$ | $a_7$ |
|---|---|---|---|---|---|---|---|
| V | 0.0116 | 0.0229 | 0.6915 | 0.00944 | 0.1372 | 0.1621 | 0.1186 |

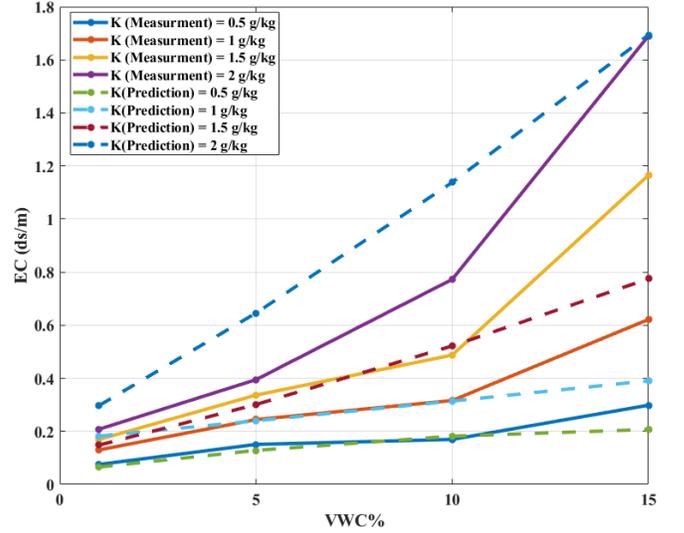

Fig. 5. Comparison of measured EC and predicted model based on Equation (11)

Another model is based on a quadratic polynomial, as shown in Equation (12) and TABLE III. The comparison between the created model and the measured values is presented in Fig. 6, demonstrating the fit of the quadratic model to the experimental data.

$$EC_w(S, \theta_w, n_e) = \\ (b_1 + b_2 S + b_3 \theta_w + b_4 S^2 + b_5 \theta_w S + b_6 \theta_w^2 + b_7 S^3 \\ + b_8 S^2 \theta_w + b_9 \theta_w^2 S) \times f_2(n_e) \quad (12)$$

TABLE III. Coefficients of Equation (12).

| Coeff | $b_1$ | $b_2$ | $b_3$ | $b_4$ | $b_5$ |
|---|---|---|---|---|---|
| Value | -0.05822 | 0.199 | 0.0049 | -0.168 | -0.0085 |
| Coeff | $b_6$ | $b_7$ | $b_8$ | $b_9$ | |
| Value | -0.000116 | 0.0444 | 0.005 | 0.00012 | |

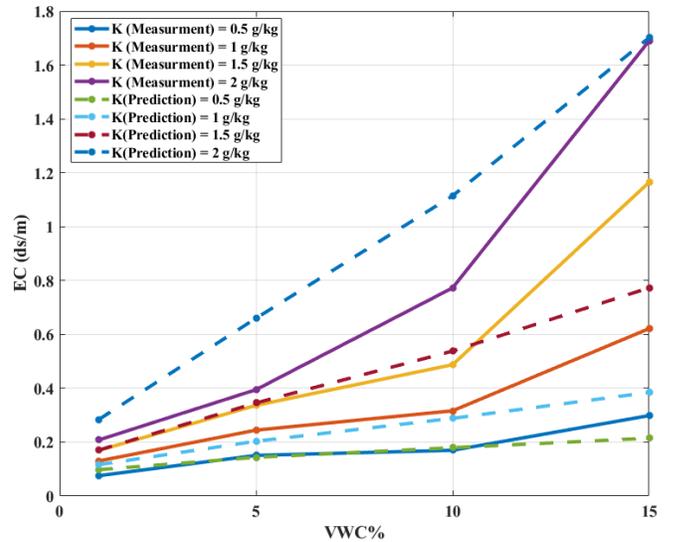

Fig. 6. Comparison of measured EC and predicted model based on Equation (12).

## B. Frequency-dependent model for salinity

Many commercial sensors measure EC at a fixed frequency, such as Troes 12 sensors operating at 70 MHz [43]. However, the models currently used for estimating soil salinity are often frequency-independent, which could lead to significant errors when measurements are made at different frequencies. Since the measurements in this study were performed across a range of frequencies, a frequency-dependent model could be proposed for more precise estimation of salinity. This is particularly important when considering the use of commercial sensors like Troes for soil salinity estimation. Such a model would provide a more accurate salinity assessment based on the sensor output, offering a practical and relatively accurate tool for real-world applications.

The frequency-dependent relationship can be defined as in Equation (13), considering variations in salinity and VWC over the frequency range, thereby improving the overall accuracy of salinity prediction compared to existing models. It should be noted that this relationship is valid for frequencies between 10-300 MHz.

$$EC_w(S, \theta_w, f, n_e) = \widehat{f_1}(S, \theta_w, f) f_2(n_e) \quad (13)$$

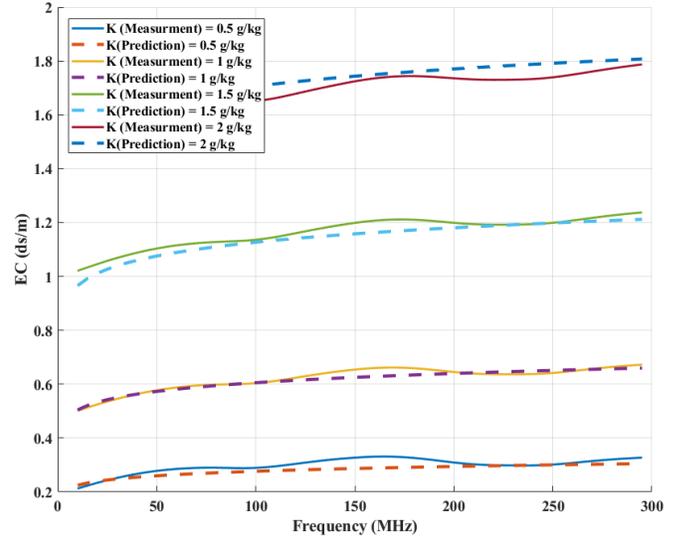

Fig. 7. Comparison of measured EC and predicted model based on Equation (14) at 15% VWC level across 5 potassium levels (0, 0.5, 1, 1.5, 2 g/kg).

The model that can be proposed for the function $\widehat{f_1}$ is a power function, with both the coefficient and exponent being dependent functions of salinity and VWC. These dependencies are represented in Equations (14), (15) and (16). The comparison between the predicted results from the model based on Equation (14) and the measured values is shown in Fig. 7.

$$\widehat{f_1}(f, S, \theta_w) = f_1(S, \theta_w) a(S, \theta_w) f^{b(S, \theta_w)} \quad (14)$$

$$a(EC_e, \theta_w) = (c_1 S^2 + c_2 S + c_3)(c_4 \theta_w^2 + c_5 \theta_w + c_6) \quad (15)$$

Where f represents frequency, measured in megahertz, and it can vary within the range of 10 to 295 MHz.

TABLE IV. Coefficients of Equation (15).

| Coeff | $c_1$ | $c_2$ | $c_3$ | $c_4$ | $c_5$ | $c_6$ |
|---|---|---|---|---|---|---|
| Value | -0.3162 | 0.7642 | 0.2068 | 0.01032 | -0.0916 | 0.7021 |

$$b(EC_e, \theta_w) = d_1 S^2 + d_2 S + d_3 S \theta_w + d_4 \theta_w^2 + d_5 \theta_w + d_6 \quad (16)$$

TABLE V. Coefficients of Equation (16).

| Coeff | $d_1$ | $d_2$ | $d_3$ | $d_4$ | $d_5$ | $d_6$ |
|---|---|---|---|---|---|---|
| Value | -0.00326 | -0.0394 | 0.001538 | -0.000795 | 0.001263 | 0.259 |

## C. Final proposed model for salinity

Soil type affects EC and salinity measurements [23]. So far, we have assumed a fixed coefficient for the soil type. To examine the impact of soil composition, a new dataset similar to the previous one, containing 20 samples, has been created. The salinity levels and VWC are the same as the previous dataset, but this dataset includes a mix of 90% sandy soil and 10% clay soil. The comparison of the measured data for this mixture with pure sandy soil is shown in Fig. 8.

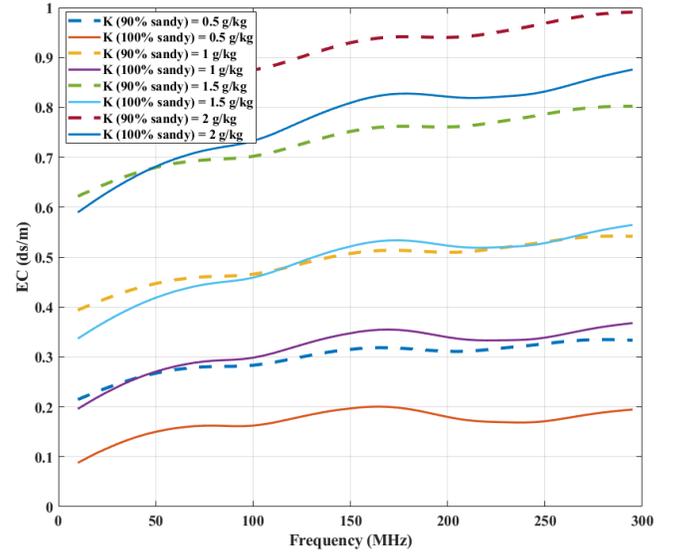

Fig. 8. Comparison of the measured EC for mixed soil (sand and clay) with pure sandy soil at 10% VWC level across 5 potassium levels (0, 0.5, 1, 1.5, 2 g/kg).

As shown in Fig. 8, a significant difference in the EC of the two soil types is observed. This test is limited in scope, and a larger dataset and more extensive testing are needed to fully understand the simultaneous effect of soil composition on moisture and salinity levels. Nevertheless, based on these initial results, we can examine the influence of two key factors: effective porosity ($n_e$) and bulk density ($\rho_b$). Since the bulk density, defined as the mass of soil per unit volume, is similar for both soil types (given that 90% of each is composed of sandy soil), we can disregard the effect of $\rho_b$ in this analysis and assume that all changes are attributed to effective porosity ($n_e$).

Effective porosity refers to the portion of the pore space in the soil that is interconnected and capable of transmitting fluids. It plays a critical role in determining ion transport pathways, which directly influence EC. Soils with higher $n_e$ provide more conductive fluid-filled pathways, resulting in higher EC values. In contrast, clay soils, which may have high total porosity, often have lower $n_e$ due to isolated pores that do not contribute to fluid flow. Thus, $n_e$ significantly affects the soil's ability to conduct electricity in its liquid phase. According to the data in [23], clay-rich soils or altered tuff materials tend to have higher overall porosity compared to sandy soils. For instance, materials like Syporex® have porosities up to 0.80, while shale or tuff particles can have porosities ranging between 0.41 and 0.64 [23]. Clean sands, however, have lower porosity, typically ranging from 0.12 to 0.35. Therefore, a functional equation for the soil's influence on conductivity is proposed in Equation (17) and TABLE VI. The results obtained from the proposed model and the measured values are shown in Fig. 9.

$$f_2(n_e) = \left(\frac{n_e}{0.3}\right)^{e_1\theta_w + e_2 S^2 + e_3 S + e_4} \quad (17)$$

TABLE VI. Coefficients of Equation (17).

| Coefficient | $e_1$ | $e_2$ | $e_3$ | $e_4$ |
|---|---|---|---|---|
| Value | -1.286 | -0.08176 | 1.613 | -2.884 |

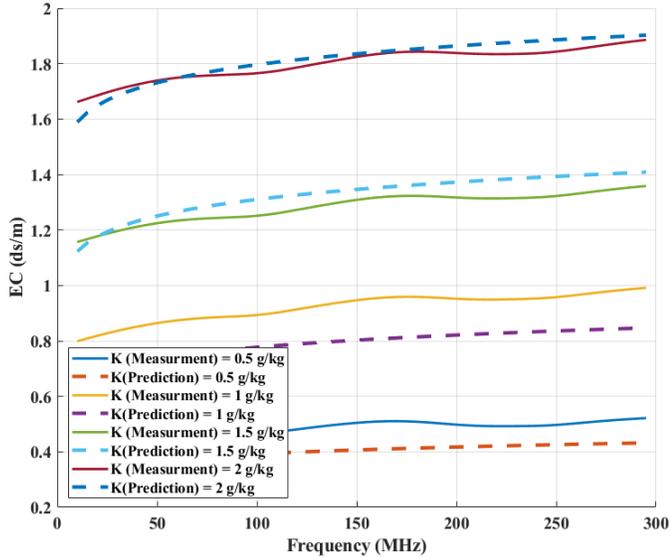

Fig. 9. Comparison of measured EC and predicted model based on Equation (17) for mixed soil at 15% VWC level across 5 potassium levels (0, 0.5, 1, 1.5, 2 g/kg).

If Equations (14), (15), (16), and (17) are substituted into Equation (13), the final equation is obtained.

## V. Conclusion

In this study, we developed a comprehensive, frequency-dependent model for predicting soil salinity based on electrical conductivity (EC) and volumetric water content (VWC). First, a diverse sandy soil dataset was constructed with varying salinity levels and VWC values. By measuring EC across a wide frequency range (10–295 MHz) and analyzing existing models, we identified key gaps in addressing the non-linear and frequency-dependent interactions among soil properties, salinity, and moisture. Initially, we proposed a frequency-independent model solely dependent on salinity and VWC. Building on this, a frequency-dependent model was developed to incorporate these interactions explicitly. Furthermore, we extended our dataset by including mixed soil types (clay and sand) to account for the effect of effective porosity. This addition allowed us to refine the model further, making it adaptable to heterogeneous soil structures. The final proposed model demonstrated consistency with measurements from commercial sensors, providing a reliable estimation of salinity. The model's flexibility and alignment with sensor data underscore its potential as a robust tool for accurately estimating soil salinity, which is essential for applications in agriculture, environmental monitoring, and soil management.

## VI. Acknowledgment

This research was supported by NTT Group (Nippon Telegraph and Telephone Corporation Group) and the Food Agility Cooperative Research Centre (CRC) Ltd, funded under the Commonwealth Government CRC Program for the 'Sustainable Sensing, Enhanced Connectivity, and Data Analytics for Precision Urban and Rural Agriculture' project at the RF and Communication Technologies (RFCT) research laboratory at the University of Technology Sydney (UTS). The CRC Program supports industry-led collaborations between industry, researchers, and the community.